\documentclass[twocolumn,showpacs,preprintnumbers,amsmath,amssymb]{revtex4}
\usepackage{epsfig}
\usepackage{graphicx}
\usepackage{dcolumn}
\usepackage{bm}

\let\jnfont=\rm
\def\NPB#1,{{\jnfont Nucl.\ Phys.\ B }{\bf #1},}
\def\PLB#1,{{\jnfont Phys.\ Lett.\ B }{\bf #1},}
\def\EPJC#1,{{\jnfont Eur.\ Phys.\ Jour.\ C }{\bf #1},}
\def\PRD#1,{{\jnfont Phys.\ Rev.\ D }{\bf #1},}
\def\PRL#1,{{\jnfont Phys.\ Rev.\ Lett.\ }{\bf #1},}
\def\MPLA#1,{{\jnfont Mod.\ Phys.\ Lett.\ A }{\bf #1},}
\def\JPG#1,{{\jnfont J.\ Phys.\ G}{\bf #1},}
\def\CTP#1,{{\jnfont Commun.\ Theor.\ Phys.\ }{\bf #1},}
\def\ZPC#1,{{\jnfont Z.\ Phys.\ C }{\bf #1},}
\def\JHEP#1,{{\jnfont JHEP \ }{\bf #1},}
\def\q_slash{\not{\hbox{\kern-2.1pt $q$}}}
\def\p_slash{\not{\hbox{\kern-4.0pt $p$}}}
\def\k_slash{\not{\hbox{\kern-2.1pt $k$}}}

\begin{document}

\preprint{\parbox{1.2in}{\noindent hep-ph/0703??? }}

\title{\ \\[10mm] Top-quark FCNC Productions at LHC
                  in Topcolor-assisted Technicolor Model}

\author{ Junjie Cao$^{1,2}$, Guoli Liu$^3$, Jin Min Yang$^4$, Huanjun Zhang$^{1,4}$ \\~~ }
\affiliation{  $^1$ Department of Physics, Henan Normal
University, Xinxiang 453007,  China \\
$^2$ Physics Department, Technion, 32000 Haifa, Israel\\
$^3$ Service de Physique Theorique CP225, Universite Libre de
          Bruxelles, 1050 Brussels, Belgium \\
$^4$ Institute of Theoretical Physics, Academia Sinica, Beijing 100080, China
 }

\begin{abstract}
We evaluate the top-quark FCNC productions induced by the
topcolor assisted technicolor  (TC2) model at the LHC. These
productions proceed respectively through the parton-level
processes $ g g \to t \bar{c}$, $c g \to t$, $c g \to t g$, $c g
\to t Z$ and $c g \to t \gamma$. We show the
dependence of the production rates on the relevant TC2 parameters
and compare the results with the predictions in the minimal
supersymmetric model. We find that for each channel the TC2 model
predicts a much larger production rate than the supersymmetric
model. All these rare productions in the TC2 model can
be enhanced above the $3\sigma$ sensitivity of the LHC. Since in
the minimal supersymmetric model only $c g \to t $ is slightly
larger than the corresponding LHC sensitivity, the observation of
these processes will favor the TC2 model over the supersymmetric
model.  In case of unobservation, the LHC can set meaningful
constraints on the TC2 parameters.

\end{abstract}
\pacs{14.65.Ha, 12.60.Fr, 12.60.Jv}

\maketitle

{\bf Introduction:~~} It is well known that flavor-changing
neutral-current (FCNC) processes have been a crucial test of the
Standard Model (SM) and a good probe for new physics beyond the
SM. As the heaviest fermion in the SM, the top quark may play a
special role in such FCNC phenomenology. In the SM the top quark
FCNC interactions are extremely suppressed
\cite{tcvh-sm} and impossible to be detected in
current and foreseeable colliders. In contrast to the SM, the new
physics models often predict much larger FCNC top quark
interactions \cite{Larios:2006pb}. Such large FCNC top quark
interactions are so far allowed by current experiments since the
Tevatron collider only gave some rather loose bounds on the FCNC
top quark decays due to the small statistics \cite{cdfd0}. The
future colliders like the LHC will allow a precision test for the
top quark properties including the FCNC interactions
\cite{Aguilar-Saavedra:2004wm}.

Once the measurement of the FCNC top quark processes is performed at
the LHC, some new physics models can be immediately tested. For
example, the FCNC top quark decays and top-charm associated
productions were found to be significantly enhanced in the minimal
supersymmetric model \cite{tcv-mssm,pptc-mssm} and technicolor
models \cite{tc-TC2,tcv-TC2}.

Although so far in the literature there are many papers
devoting to the new physics predictions for the FCNC top quark productions
at the LHC, usually different processes are treated in different papers.
Since these FCNC processes are correlated in a given new physics model,
it is necessary to give a comprehensive study of all these processes in
one paper. Recently, such an effort was given for the popular supersymmetric
models \cite{cao-pp2tc}. In this work we perform  a comprehensive analysis
for the FCNC top quark productions in TC2 model.
We will consider the production channels
\begin{eqnarray}
g g \to t \bar{c},  ~c g \to t,  ~c g \to t g,
~c g \to t Z, ~c g \to t \gamma  \label{pro-6}
\end{eqnarray}
Some of these processes have been studied in the literature: $gg
\to t\bar c$ was studied in the second and third papers in
\cite{tc-TC2}, but only the $s$-channel contributions were
considered; $cg\to tV$ ($V$ is a vector boson) were studied in
\cite{tztrtg}, but all box diagrams were ignored. The other
process $cg \to t$ in TC2 model has not been studied in the
literature. In this work we consider all these productions and
compare their rates. Also, we will compare the TC2 results with
the predictions of supersymmetric models. Note that in our studies
the parton-level processes will be used to label the corresponding
hadronic productions and the charge-conjugate channel for each
production is also included. \vspace*{0.5cm}

 {\bf About TC2 Model:} Before our calculations we recapitulate
the basics of TC2 model. As is well known, the fancy idea of
technicolor tries to provide an elegant dynamical mechanism for
electroweak symmetry breaking, but it encounters great difficulty
when trying to generate fermion masses, especially the heavy top
quark mass. The TC2 model \cite{TC2} combines technicolor
interaction with topcolor interaction, with the former being
responsible for electroweak symmetry breaking and the latter for
generating large top quark mass. This model so far survives current
experimental constraints and remains one of the candidates of new
physics.

The TC2 model predicts a number of pseudo-Goldstone bosons like
the top-pions ($ \pi^0_t$ and $ \pi^\pm_t$) at the weak scale
\cite{TC2}. The top quark
interactions are altered with respect to the SM predictions since
it is treated differently from other fermions in TC2 model. For
example, the TC2 model predicts some anomalous couplings for the
top quark, such as the tree-level FCNC coupling $t \bar c \pi^0_t$
and the charged-current $t \bar b \pi^-_t$ coupling given by
\begin{eqnarray}
& &\frac{(1 - \epsilon ) m_{t}}{\sqrt{2}F_{t}}
     \frac{\sqrt{v^2-F_{t}^{2}}}{v} \left (
 i K_{UL}^{tt}K_{UR}^{tt } \bar{t}_L t_{R} \pi_t^0  \right . \nonumber \\
&&  + \sqrt{2}K_{UR}^{tt} K_{DL}^{bb}\bar{t}_R b_{L} \pi_t^-
    + i K_{UL}^{tt} K_{UR}^{tc} \bar{t}_L c_{R} \pi_t^0  \nonumber \\
&& + \sqrt{2} K_{UR}^{tc} K_{DL}^{bb} \bar{c}_R b_{L}\pi_t^-
           + K_{UL}^{tt} K_{UR}^{tt } \bar{t}_L t_{R} h_t^0   \nonumber \\
&& \left. + K_{UL}^{tt} K_{UR}^{tc} \bar{t}_L c_{R} h_t^0 + h.c.  \right ) ,
\label{FCNH}
\end{eqnarray}
where the factor $\sqrt{v^2-F_t^2}/v$ ( $v \simeq 174$ GeV )
reflects the effect of the mixing between the top-pions and the
would-be Goldstone bosons \cite{9702265}. $K_{UL}$, $K_{DL}$ and
$K_{UR}$ are the rotation matrices that transform respectively the
weak eigenstates of left-handed up-type, down-type and right-handed
up-type quarks to their mass eigenstates, whose values can be
parameterized as \cite{tc-TC2}
\begin{eqnarray}
&&  K_{UL}^{tt} \simeq K_{DL}^{bb} \simeq 1,
    ~~K_{UR}^{tt}\simeq \frac{m_t^\prime}{m_t} = 1-\epsilon, \\
&&  K_{UR}^{tc}\leq \sqrt{1-(K_{UR}^{tt})^2}
    =\sqrt{2\epsilon-\epsilon^{2}}, \label{FCSI}
\end{eqnarray}
with $m_t^\prime$ denoting the topcolor contribution to the top
quark mass. In Eq.(\ref{FCNH}) we neglected the mixing between up
quark and top quark. Note that in TC2 model a CP-even scalar called
top-Higgs ($h^0_t$) may also exist, whose couplings are similar to
the neutral top-pion\cite{tc-TC2}.

The parameters involved in our calculations are: the masses of the
top-pions and top-Higgs, the parameter $K_{UR}^{tc}$, the top-pion
decay constant $F_t$ and the parameter $\epsilon$, which
parameterizes the portion of the extended-technicolor contribution
to the top quark mass. In our study we take $m_t=180.7$ GeV
\cite{0612061} and $F_t=50$ GeV. Since the mass splitting between
neutral and charged top-pion is very small, we assume
$m_{\pi}^0=m_{\pi}^{\pm}$. The top-pions mass is model-dependent and
is usually of a few hundred GeV \cite{TC2}. About the top-Higgs
mass, ref. \cite{tc-TC2} gave a lower bound of  about $2m_t$, but it
is an approximate analysis and the mass below $t\bar t$ threshold is
also possible \cite{9809470}. In our analysis we assume
\begin{equation}
m_{\pi^0_t}=m_{\pi^\pm_t}=m_{h^0_t}\equiv M_{TC}.
\end{equation}
\vspace*{0.5cm}

\begin{figure}[bt]
\epsfig{file=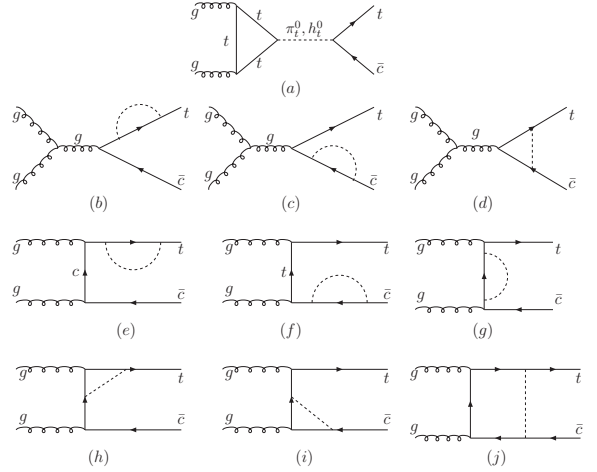,width=8cm} \caption{Feynman diagrams for $g g
\to t \bar{c}$ in TC2 model. The boson in each loop denotes a
neutral top-pion, top-Higgs or a charged top-pion, while the
fermion in each loop can be a top or bottom quark depending on the
involved boson being neutral or charged. } \label{fig1}
\end{figure}

\begin{figure}[bt]
\epsfig{file=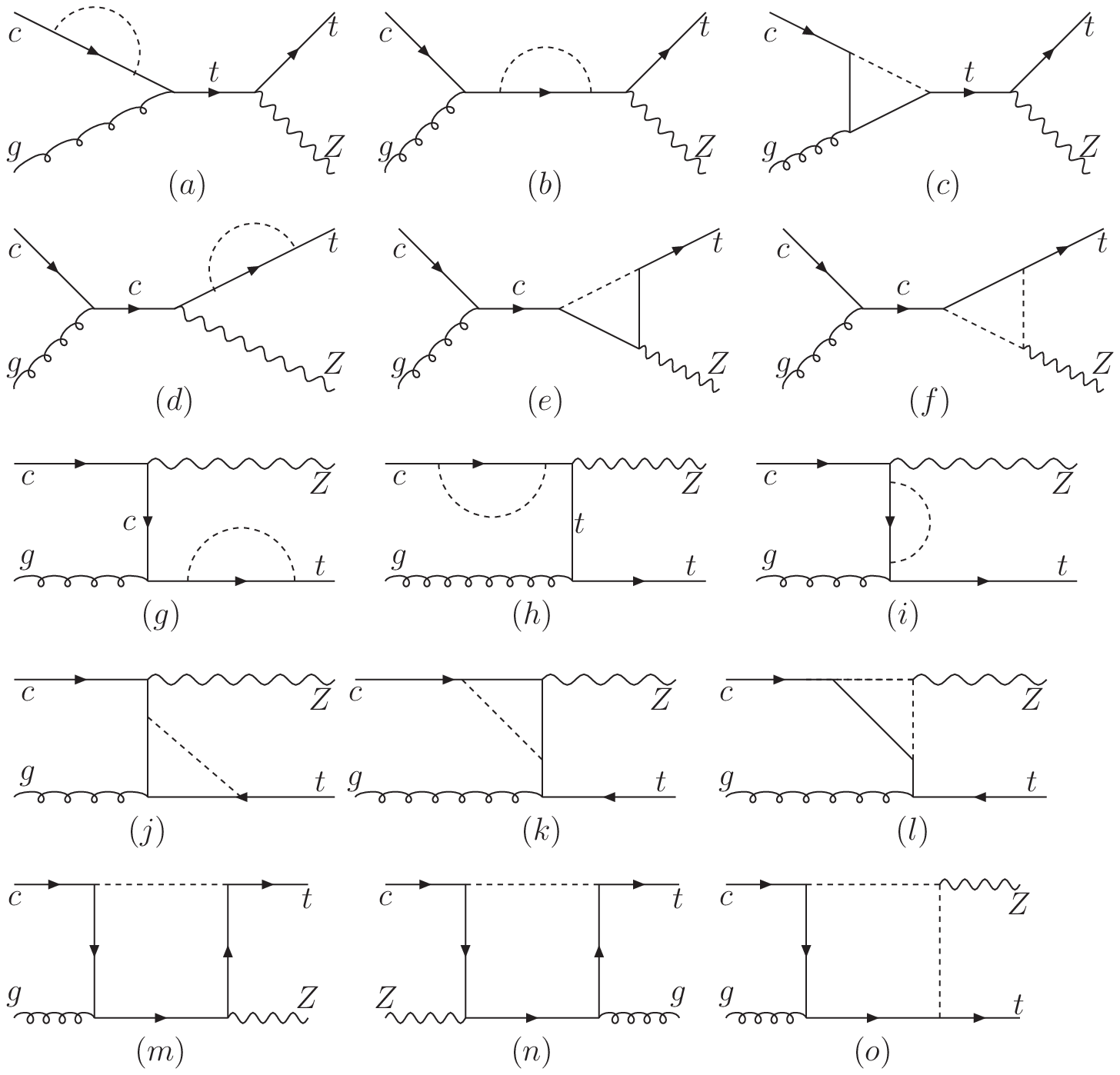,width=8cm}
\caption{Feynman diagrams for $c g \to t Z$ in TC2 model.
      The boson in each loop denotes a neutral top-pion, top-Higgs or
      a charged top-pion, while the fermion in each loop can be a top
      or bottom quark depending on the involved boson being neutral or
      charged.} \label{fig2}
\end{figure}

 {\bf Calculations:~~} For the parton-level processes
in eq.(\ref{pro-6}) we only plot the Feynman diagrams in
Figs.~\ref{fig1} and \ref{fig2} for $gg \to t \bar c$ and $cg \to
tZ$, respectively. Other processes have similar Feynman diagrams
which can be easily obtained from Figs.~\ref{fig1} and \ref{fig2}.
For example, $cg \to tg $ can be straightforwardly obtained from
Fig.~\ref{fig1}, and $cg \to t\gamma$ can be obtained from
Fig.\ref{fig2} by removing some diagrams with non-exist vertices.
The calculations for these production processes are
straightforward. Here we take the calculation of $gg\to t\bar c$
as an example. Its amplitude takes the form
\begin{eqnarray}
&& \frac{g_s^2m_t^2}{2
F_t^2}\frac{v^2-F_t^2}{v^2}K_{UR}^{tt}K_{UR}^{tc}
\epsilon_\mu(k_1)\epsilon_\nu(k_2) \nonumber \\
&& \times \sum_iT^i{\bar{u}(p_t) \Gamma_i^{\mu\nu} P_R v(p_c)},
\label{amp}
\end{eqnarray}
where the sum is over all the Feynman diagrams in Fig.~\ref{fig1},
$T_i$ are color factors, $P_R=(1+\gamma_{5})/2$,
$k_{1,2}$ denote the momentum of two incoming gluons and
$p_{t,c}$ the momentum of outgoing top and anti-charm quarks,
and $\Gamma_i^{\mu\nu}$ is given by
\begin{eqnarray}
&& c^i_1
p_t^{\mu}p_t^{\nu}+c^i_2p_c^{\mu}p_c^{\nu}+c^i_3p_t^{\mu}p_c^{\nu}
   +c^i_4p_t^{\nu}p_c^{\mu} +c^i_5p_t^{\mu}\gamma^\nu
   +c^i_6p_c^{\mu}\gamma^\nu \nonumber\\
&&
   +c^i_7p_c^{\nu}\gamma^\mu
   +c^i_8p_t^{\nu}\gamma^\mu
   +c^i_9g^{\mu\nu}
   +c^i_{10}\gamma^{\mu}\gamma^{\nu}
   +c^i_{11}p_t^{\mu}p_t^{\nu}{\not{k}}_2 \nonumber\\
&&
  +c^i_{12}p_c^{\mu}p_c^{\nu}{\not{k}}_2
   +c^i_{13}p_t^{\mu}p_c^{\nu} {\not{k}}_2
   +c^i_{14}p_t^{\nu}p_c^{\mu}{\not{k}}_2
    \nonumber\\
&& +c^i_{15}p_t^{\mu}\gamma^\nu{\not{k}}_2
+c^i_{16}p_c^{\mu}\gamma^\nu{\not{k}}_2
+c^i_{17}p_c^{\nu}\gamma^\mu {\not{k}}_2 \nonumber\\
&&
   +c^i_{18}p_t^{\nu}\gamma^\mu {\not{k}}_2
   +c^i_{19}g^{\mu\nu} {\not{k}}_2
   +c^i_{20}i\varepsilon^{\mu\nu\alpha\beta}\gamma_\alpha k_{2\beta}.
\label{m-e}
\end{eqnarray}
Here the coefficients $c^i_j$ are obtained by the straightforward
calculation of each Feynman diagram in Fig.~\ref{fig1}, which are
composed of scalar loop functions \cite{Hooft} and can be calculated
by using LoopTools \cite{Hahn}. The calculations of the loop
diagrams are tedious and the analytical expressions for the
coefficients $c^i_j$ are lengthy, so we do not present the explicit
expressions of $c_i$s here.

The hadronic cross section at the LHC is obtained by convoluting the
parton cross section with the parton distribution functions. In our
calculations we use CTEQ6L \cite{cteq} to generate the parton
distributions with the renormalization scale $\mu_R $ and the
factorization scale $\mu_F$ chosen to be $\mu_R = \mu_F = m_t$. To
make our predictions more realistic, we applied some kinematic cuts.
For example, we require that the transverse momentum of each
produced particle larger than 15 GeV and its pseudo rapidity less
than 2.5 in the laboratory frame. For $c g \to t$ followed by $t\to
W b$, we do not require the top quark exactly on mass shell and
instead we require the invariant mass of bottom quark and $W$-boson
in a region of $m_t - 3 \Gamma_t \leq M_{b W} \leq m_t + 3 \Gamma_t
$ ($\Gamma_t$ is the top quark width). This requirement was used in
\cite{Hosch} to investigate the observability of this channel at
hadron colliders in the effective Lagrangian framework.
\vspace*{0.5cm}

\begin{figure}[bt]
\epsfig{file=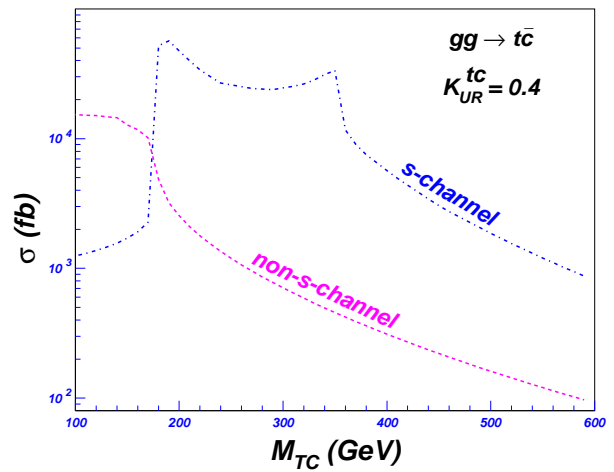,width=8cm}
\caption{ The hadronic cross section of the production proceeding
          through $g g \to t \bar{c}$ versus $M_{TC}$.}
\label{fig3}
\end{figure}

{\bf Numerical results:~~} Since the cross section for each
channel is simply proportional to $K_{UR}^{tc}$ as shown in
eq.(\ref{amp}), here we do not show the dependence on
$K_{UR}^{tc}$. We will fix $K_{UR}^{tc} = 0.4$ and show the
dependence on $M_{TC}$. In Fig. \ref{fig3} we show the hadronic
cross section of the production proceeding by the parton-level
process $gg\to t \bar c$ versus $M_{TC}$, where the $s$-channel
and non-$s$-channel contributions are shown separately.
>From Fig. \ref{fig3} we see that the contributions are dominated by
the $s$-channel process for a heavy top-pion and
there exist three regions of $M_{TC}$.
In the range $m_t < M_{TC} < 2 m_t$, the cross section is maximal
and can reach about $30$ pb.
The reason is that in this region $t\bar{c}$ is the dominant decay mode
of $\pi^0_t$ and $h^0_t$ which can be produced on-shell through the
$s$-channel. When $M_{TC}$ passes the threshold of $2 m_t$ and
keeps increasing, the cross section drops quickly since
the $t\bar t$ is becoming the dominant decay mode of $\pi^0_t$
and $h^0_t$.  In the light mass region $M_{TC} < m_t$, which is
disfavored by $R_b$ \cite{kuang}, the non-$s$-channel
contributions are dominant and the $s$-channel contributions are
suppressed since the top-pion and the top-Higgs in the $s$-channel
cannot be produced on-shell.

\begin{figure}[hbt]
\epsfig{file=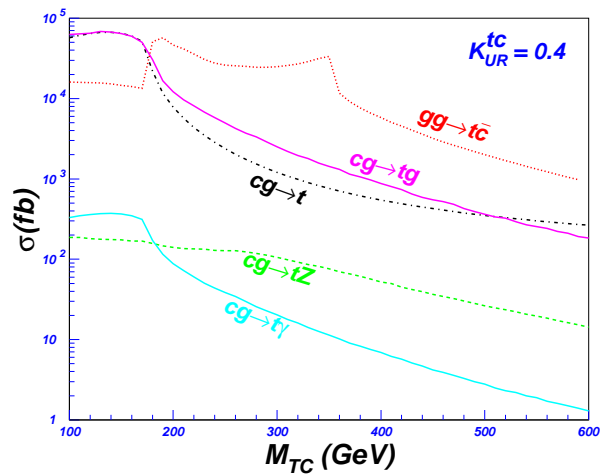,width=8cm} \caption{ The hadronic cross
sections of the productions proceeding through the parton-level
processes labelled on each curve.} \label{fig4}
\end{figure}

The total hadronic cross sections for all these processes
are plotted in Fig.\ref{fig4} for comparison.
We see that the production proceeding through
$gg\to t\bar c $ has the largest rate for a heavy top-pion.
Of course, the productions in the two channels of $gg\to t\bar c $
and  $cg\to tg $ cannot be distinguished from each other since
the charm quark jet cannot be distinguished from the gluon jet.
Therefore, the cross sections of these  two channels should be
summed, which gives a signal of an energetic lepton (electron or muon)
plus two jets (one of them is $b$-jet) plus missing energy.

Now we compare the TC2 predictions with the predictions in
supersymmetric model in Table 1. The TC2 predictions are taken
from Fig.\ref{fig4} for $M_{TC}=300$ GeV and $K_{UR}^{tc}=0.4$,
while the predictions of the minimal supersymmetric model are the
maximal values taken from \cite{cao-pp2tc}. The parameters
$\delta_{LL}$ and $\delta_{LR}$ parameterize the mixing between
top-squarks and charm-squarks and their definitions can be found
in \cite{cao-pp2tc}. We see that for each channel the TC2 predicts
a much larger production rate than the minimal supersymmetric
model. \vspace*{0.2cm}

\noindent {\small Table 1: The hadronic cross sections of
top-quark FCNC productions in TC2 and the minimal supersymmetric
model. The TC2 predictions are taken from Fig.\ref{fig4}
for $M_{TC}=300$ GeV and $K_{UR}^{tc}=0.4$, while the
predictions of the minimal supersymmetric model are the maximal
values taken from \cite{cao-pp2tc}.
The corresponding charge-conjugate channels are also
included. The LHC sensitivities in the last column are for 100
fb$^{-1}$ integrated luminosity.} \vspace*{-0.3cm}

\begin{center}
\begin{tabular}{|l|l|l|l|l|} \hline
        & \multicolumn{2}{c|} {SUSY}  & ~~TC2~~ &    ~~LHC sensitivity~~ \\ \cline{2-3}
                  & $\delta_{LL}\neq 0$ & $\delta_{LR}\neq 0$ &  & ~~~~~~~~~~$3 \sigma$ \\ \hline
$gg \to t\bar{c}$ & 240 fb & 700 fb & 30 pb& 1500 fb
\cite{pptc-Han,pptc-background}\\ \hline $cg \to t$        & 225 fb
& 950 fb & 1.5 pb & 800 fb \cite{Hosch} \\\hline $ c g \to tg$     &
85 fb  & 520 fb & 3 pb & 1500 fb \cite{pptc-Han,pptc-background}\\
\hline $cg \to t \gamma$ & 0.4 fb & 1.8 fb & 20 fb   & 5 fb
\cite{zt-Aguilar} \\ \hline $cg \to tZ$       & 1.5 fb & 5.7 fb &
100 fb   & 35 fb \cite{zt-Aguilar}  \\ \hline
\end{tabular}
\end{center}
\vspace*{0.2cm}

In Table 1 we also list the LHC sensitivity with 100 fb$^{-1}$
integrated luminosity.  Such sensitivity for each production channel
has been intensively investigated in the literature listed in Table
1. Although these sensitivities are based on the effective
Lagrangian approach and may be not perfectly applicable to a
specified model, we can take them as a rough criteria to estimate
the observability of these channels. Comparing these sensitivities
with the TC2 predictions, we see that all the productions can be
above the $3\sigma$ sensitivity of the LHC for the chosen TC
parameters. But for the minimal supersymmetric model, only the
prediction for $c g \to t $ is slightly larger than the
corresponding LHC sensitivity. Therefore, if these rare processes
are observed at the LHC, the TC2 model, rather than supersymmetry,
will be favored. Of course, in case of unobservation of these rare
productions, the LHC can set meaningful constraints on TC2
parameters.

Note that in Table 1 we did not list the SM predictions, which have
not been calculated in the literature since they must be far below
the observable level due to the extremely suppressed top quark FCNC
interactions \cite{tcvh-sm}. Also, we did not list the comparison of
TC2 and supersymmetry predictions for various top quark FCNC decays,
which can be found in the last reference of \cite{tc-TC2}. From
there we wee that for top quark FCNC decays the TC2 predictions are
also much larger than supersymmetry predictions.  So the potentially
large top quark FCNC interaction is one characteristic of TC2 model
and will serve as a crucial test for this model at future collider
experiments.
 \vspace*{0.4cm}

{\bf Conclusions:~~} We evaluated the top-quark FCNC productions
in the top-color assisted technicolor model at the LHC. These
productions proceed respectively by the parton-level processes $ g
g \to t \bar{c}$, $c g \to t$, $c g \to t g$, $c g \to t Z$ and $c
g \to t \gamma$. We found that the predictions of the production
rates in this model are much larger than in the supersymmetric
model and all the productions can be enhanced above the $3\sigma$
sensitivity of the LHC. Since in the minimal supersymmetric model
only $c g \to t $ is slightly larger than the corresponding LHC
sensitivity, the observation of these processes will imply that
the TC2 model is more favored than the supersymmetric model.  In
case of unobservation of these rare productions, the LHC can set
meaningful constraints on TC2 parameters. \vspace*{0.2cm}

{\bf Acknowledgment:} JMY thanks K. Hikasa for helpful discussions
and acknowledges the COE program of Japan for supporting a visit
in Tohoku University where part of this work is done.
This work is supported in part by a
fellowship from the Lady Davis Foundation at the Technion, by the
Israel Science Foundation (ISF),  the National Natural Science
Foundation of China under Grant No. 10475107 and 10505007, and by
the IISN and the Belgian science policy office (IAP V/27).

\end{document}